\documentclass[twocolumn,floatfix,superscriptaddress,citeautoscript]{revtex4}

\usepackage{graphicx}

\begin{document}

\title{Picovoltmeter for probing vortex dynamics in a single weak-pinning Corbino channel}

\author{T.W. Heitmann}
\affiliation{Department of Physics, Syracuse University, Syracuse, NY 13244-1130}
\author{K. Yu}
\affiliation{Department of Physics, Syracuse University, Syracuse, NY 13244-1130}
\author{C. Song}
\affiliation{Department of Physics, Syracuse University, Syracuse, NY 13244-1130}
\author{M.P. DeFeo}
\affiliation{Department of Physics, Syracuse University, Syracuse, NY 13244-1130}
\author{B.L.T. Plourde}
\email[]{bplourde@phy.syr.edu}
\affiliation{Department of Physics, Syracuse University, Syracuse, NY 13244-1130}
\author{M.B.S. Hesselberth}
\affiliation{Kamerlingh Onnes Laboratorium, Leiden University, P.O. Box 9504, 2300 RA Leiden, The Netherlands}
\author{P.H. Kes}
\affiliation{Kamerlingh Onnes Laboratorium, Leiden University, P.O. Box 9504, 2300 RA Leiden, The Netherlands}

\date{\today}

\begin{abstract}
We have developed a picovoltmeter using a Nb dc Superconducting QUantum Interference Device (SQUID) for measuring the flux-flow voltage from a small number of vortices moving through a submicron weak-pinning superconducting channel. We have applied this picovoltmeter to measure the vortex response in a single channel arranged in a circle on a Corbino disk geometry. The circular channel allows the vortices to follow closed orbits without encountering any sample edges, thus eliminating the influence of entry barriers.
\end{abstract}

\maketitle

\section{Introduction}
The dynamics of vortices in confined superconductor geometries has generated much interest in recent years, with studies of both fundamental properties of vortex matter as well as devices based on the motion of vortices.
Nanoscale channels for guiding vortices through superconducting films with a minimal influence from pinning have been developed for explorations of vortex melting \cite{besseling03}, commensurability \cite{pruymboom88}, mode locking \cite{kokubo02}, and ratchets \cite{yu07}.
These channels are typically arranged across the width of a superconducting strip, so that the vortices enter the channel at one edge of the strip and exit
at the other edge, 
resulting in edge barriers to the vortex motion through the channels \cite{vodolazov00, plourde01,benkraouda98}.
The strip geometry also allows for the use of multiple channel copies in parallel to boost the flux-flow signal strength for measurement with a room-temperature amplifier. 

It is possible to eliminate the edge barriers characteristic of a strip arrangement by using a Corbino geometry, consisting of a superconducting disk, with current injected radially between the center and the perimeter. Vortices in such a disk experience an azimuthal Lorentz force and can flow in closed circular orbits without crossing any edges. The Corbino geometry has been used for many studies of vortex matter in different superconductors, including bulk crystals of YBCO \cite{lopez99} and NbSe$_2$ \cite{paltiel00}. 
We have patterned thin-film Corbino disks with submicron circular channels for probing vortex dynamics in a narrow region free from edge barriers.
Because the current density decreases radially in a Corbino geometry, such that vortices at different radii experience a different Lorentz force \cite{babaei01}, we have designed our devices to have only a single channel. This poses a challenge for the amplifier used to detect the vortex motion.
In this article, we present a scheme for driving a small number of vortices through a single, circular submicron channel and a picovoltmeter for resolving the ensuing flux-flow voltages.

\section{Channel fabrication}
\begin{figure}[htb]
\centering
  \includegraphics{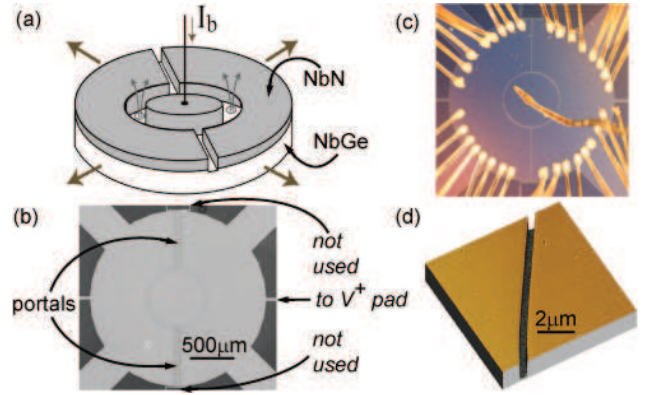}
  \caption{(Color online) (a) Corbino channel schematic. (b) Scanning electron micrograph of Corbino disk; narrow lines at top and bottom not connected to disk. (c) Optical micrograph of disk showing wirebonds. (d) Atomic Force Microscope (AFM) image of channel.
\label{fig:schematic}}
\end{figure}

We fabricate our channels from bilayers of $200$~nm-thick films of amorphous-NbGe, an extremely weak-pinning superconductor ($T_c^{NbGe}=2.88$~K), and $50$~nm-thick films of NbN, with relatively strong pinning ($T_c^{NbN}=9.6$~K), on a Si substrate. 
After patterning and etching a $1.5$~mm-diameter Corbino disk into such a bilayer, we define a $520$~nm-wide channel in a circle with a $500$~$\mu$m diameter using electron beam lithography (Fig. \ref{fig:schematic}). We etch this region down to a depth of $120$~nm using a reactive ion etch with CF$_4$, thus completely removing the NbN in this region and etching partially into the NbGe layer.
In addition to the circular channel, we etch two radial channels (portals) that extend from opposite sides of the circular channel out to the edge of the Corbino disk. These portals allow for the introduction of vortices into the channel by field-cooling from temperatures $T_c^{NbGe} < T < T_c^{NbN}$, as they break up the circulating supercurrent in the outer NbN region. 

We attach leads for driving the bias current $I_b$ through the Corbino disk using $1.25$~mil Al wirebonds, with $32$ $I_b^+$ bonds around the perimeter of the disk.
The $I_b^-$ connection to the center 
consists of a superconducting Nb wirebond using annealed $2$~mil Nb wire 
[Fig. \ref{fig:schematic}(c)].
The azimuthal Lorentz force from $I_b$ causes the vortices to flow around the channel when this force exceeds the residual pinning in the NbGe channel.
The ensuing vortex dynamics in the channel can be characterized by measuring the radial voltage drop across the channel, $V_x$, which is proportional to the vortex velocity and density.

\section{Picovoltmeter design and characterization}
The flux-flow voltage for vortices moving in a single channel at a low velocity can be quite small. For example, vortices with a density corresponding to a magnetic induction of $B_{ch}=1$~G in the 
channel moving at a velocity of $1$~m/s, less than one percent of the typical Larkin-Ovchinnikov instability velocity for NbGe \cite{larkin75, babic04}, produce a flux-flow voltage of $\sim$ ~$50$~pV.
In order to resolve such signals, we have developed a voltmeter, based on a Nb dc SQUID, 
which we obtained from ez SQUID. 
Sensitive voltmeters were one of the original applications of SQUIDs \cite{clarke66}, and SQUID voltmeters have been used previously to probe the nature of the vortex state in bulk crystals of YBCO \cite{gammel91} and BSCCO \cite{safar92}. To the best of our knowledge, our scheme is the first application of a dc SQUID to form a picovoltmeter for resolving vortex dynamics in a patterned thin-film structure, as well as in a Corbino geometry.

\begin{figure}[htb]
\centering
  \includegraphics{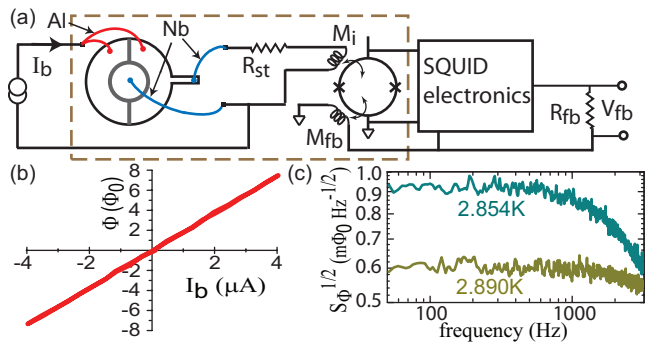}
  \caption{(Color online) (a) Voltmeter circuit schematic; portions outside dashed box are at room temperature. (b) Flux coupled to SQUID vs. $I_b$ at $4.2$~K. (c) Flux noise spectra above and below $T_c^{NbGe}$, both measured with $I_b=0$, $H_a=0$.
\label{fig:SQUID}}
\end{figure}

We connect the voltage leads across the NbGe channel to the SQUID input coil with a resistor, $R_{st}$, consisting of a segment of brass foil ($3.7$ x $3.2$ x $0.025$~mm$^3$). 
This converts the flux-flow voltage to a current through the input coil, which has a self inductance $L_i$ and a mutual inductance $M_i$ to the SQUID [Fig. \ref{fig:SQUID}(a)]. Except for $R_{st}$, all of the voltage connections are superconducting. We make the voltage contacts on the Corbino disk with Nb wirebonds, where the $V_x^-$ connection shares the superconducting Nb wire to the center of the Corbino disk with the $I_b^-$ connection, while the $V_x^+$ Nb wirebond is attached to a pad of the NbGe/NbN bilayer that extends from the perimeter of the Corbino disk [Fig. \ref{fig:schematic}(b)]. 
The other ends of these Nb wirebonds are attached to superconducting solder-tinned copper traces on our chip carrier using Pb washers. 
The $V_x^+$ connection is soldered to $R_{st}$, then the 
traces are attached to a twisted pair of $3$~mil Nb wire with a second set of screw terminals using Pb washers. Finally, this Nb twisted pair is coupled to the input circuit on the SQUID holder with superconducting screw terminals.

We operate the SQUID in a conventional flux-locked loop, using a $4$~MHz electronics system from ez SQUID, with the feedback signal $V_{fb}$ supplied through a feedback resistor $R_{fb}$ to a wire-wound coil 
with a mutual inductance $M_{fb}$ to the SQUID. The SQUID holder is mounted in a Nb cylindrical shield that is closed on one end. 
The entire bottom end of the experimental insert is enclosed in a Pb cylindrical shield, and the dewar is surrounded by a $\mu$-metal shield that is closed on the bottom. 

A simple circuit analysis leads to the following expression relating $V_{fb}$ to the voltage across the channel $V_x$:
\begin{equation}
V_{fb} = \left(\frac{R_{fb}}{R_{st}}\right) \left(\frac{M_i}{M_{fb}}\right) V_x.
\label{eq:gain}
\end{equation}
The ratio $R_{fb}/M_{fb}$ can be obtained in the usual way by measuring the difference in $V_{fb}$ with the SQUID locked in adjacent wells ($590$~mV), and $M_i$ was measured to be $6.6$~nH through a separate calibration.
We estimate $R_{st}$ to be $2$~m$\Omega$ based on the size of the brass foil, but we can obtain a more careful calibration of the voltmeter gain through a series of low-temperature measurements.
We first measure the current-voltage characteristic of the Corbino channel at $4.2$~K, where the NbGe channel is in the normal state, while the NbN banks are superconducting. Figure \ref{fig:SQUID}(b) shows the flux coupled to the SQUID  
plotted against the bias current $I_b$, which was monitored by tracking the voltage drop across a room-temperature current-sensing resistor.
Because the channel is in the normal state, $I_b$ divides between the channel, with resistance $R_n$, and the SQUID input coil with series resistance $R_{st}$. Based on $4.2$~K measurements of similar NbGe channels of various geometries, we estimate $R_n$ for this Corbino disk to be 
$3$~m$\Omega$.

The flux noise at $I_b=0$ and zero field is essentially white with a high-frequency roll-off determined by $L_i$ and the relevant resistance in the voltmeter circuit. 
By comparing the flux noise below this roll-off for temperatures above and below $T_c^{NbGe}$, we can obtain measurements of both $R_{st}$ and $R_n$. Above $T_c^{NbGe}$, the flux noise is determined by the Nyquist noise current generated by $R_{st}+R_n$ flowing through $M_i$, while for $T < T_c^{NbGe}$, the only resistance is $R_{st}$ [Fig. \ref{fig:SQUID}(c)]. 
In both cases the flux noise is more than an order of magnitude larger than the intrinsic flux noise for the SQUID, $10$~$\mu \Phi_0/Hz^{1/2}$ at $100$~Hz and $4.2$~K from a separate measurement with the voltmeter circuit disconnected from the input coil. 
This analysis leads to $R_{st}=1.9$~m$\Omega$ and $R_n=2.6$~m$\Omega$, consistent with our rough estimates. 
In addition, the location of the flux noise roll-off corresponds to $L_i=110$~nH, consistent with the design of the input coil on this particular SQUID. 

Applying our measured circuit and SQUID parameters with Eq. \ref{eq:gain} yields a gain of $9.8 \times 10^8$. The measured flux noise at $T=2.854$~K can be referred back as a voltage noise across the Corbino channel of $0.55$~pV/Hz$^{1/2}$. Integrating over the $2.7$~kHz bandwidth of the voltmeter yields an rms noise level of $25$~pV. 
Of course, this noise could be reduced further, simply by using a smaller value of $R_{st}$, with a concomitant reduction in the measurement bandwidth.

\section{Measurements of Flux-Flow in Corbino Channel with Picovoltmeter}
\begin{figure}[htb]
\centering
  \includegraphics{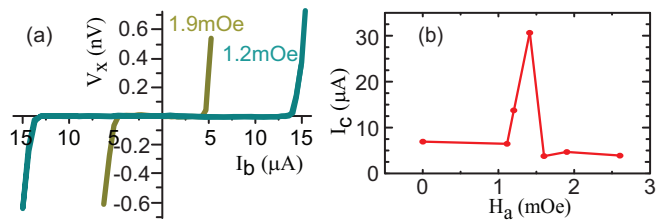}
  \caption{(Color online) (a) IVCs for two different cooling fields at $T=2.874$~K. (b) $I_c$ for different $H_a$, extracted from IVCs.  
\label{fig:IVCs}}
\end{figure}

We have applied our voltmeter to measure the current-voltage characteristics (IVCs) of our single Corbino channel at $T=2.874$~K for several different cooling fields $H_a$. 
During our measurements, the Corbino disk and SQUID circuitry are immersed in a pumped helium bath.
For each value of $H_a$, the insert was raised just above the bath and heated to $6$~K, above $T_c^{NbGe}$, and was then cooled in $H_a$, generated with a superconducting coil.
We recorded $V_{fb}$ and 
$I_b$ with a digital oscilloscope, taking $1024$ averages per point.  
The IVCs [Fig. \ref{fig:IVCs}(a)] exhibit a zero-voltage region at small $I_b$ followed by an increasing flux-flow voltage for $I_b$ beyond a depinning critical current $I_c$. 
Using a voltage criterion of $50$~pV, we extract $I_c(H_a)$, which has a peak around $1.4$~mOe. 
This peak points to zero absolute field, or the limit of no vortices trapped in the channel. In contrast to a superconducting strip, which has a moderate critical current even in zero applied field corresponding to the entry of vortices and anti-vortices at the strip edges from the self-field of the strip \cite{vodolazov00, benkraouda98}, the Corbino disk should have a large critical current when no vortices are present, as $I_c$ in this case would correspond to the breakdown of superconductivity in the entire disk.

While the values of $H_a$ in our measurements are rather small, the intermediate field-cooling scheme generates substantial flux-focusing effects into the NbGe channel due to the superconducting NbN.
A rough estimate, considering screening currents around the central disk of NbN inside the channel, and along the outer NbN banks 
\cite{Clem94, vodolazov00}, indicates the enhancement of 
$B_{ch}$ may be of the order of $10$~x~$H_a$. 
Thus, cooling in an absolute applied field of $1$~mOe should nucleate at least $\sim10$ vortices in the channel, where the effective area includes not only the channel, but the penetration of the vortex circulating currents into the NbN banks as well.
A more detailed treatment of this flux focusing in the channels is beyond the scope of this paper.

Our measurements presented here have been performed close to $T_c^{NbGe}$, where the 
residual pinning in the channels was especially weak.
This was necessary because of our wiring configuration, where the $I_b^-$ lead was shared with the $V_x^-$ connection to the center of the Corbino disk along a superconducting Nb wirebond. This Nb wirebond had a somewhat small critical current, thus requiring us to operate at temperatures where 
$I_c$ was below the Nb wirebond critical current. In future Corbino measurements, it should be possible to separate these leads attached to the center of the disk, with the Nb wirebond used only for the SQUID input connection, which does not need to sustain large currents because of the presence of $R_{st}$, with separate Al wirebonds for the $I_b^-$ connection, which does not need to be superconducting.

In summary, we have demonstrated a SQUID picovoltmeter circuit for resolving vortex dynamics in a single, weak-pinning superconducting channel in a Corbino geometry. This technique will be useful for investigations of small numbers of vortices moving at low velocities in nanofabricated structures, such as vortex ratchets.

\section*{ACKNOWLEDGMENTS}
We thank R. McDermott and M. M\"uck for stimulating discussions. 
This work was supported by the National Science Foundation under Grant DMR-0547147. We acknowledge use of the Cornell NanoScale Facility, a member of the National Nanotechnology Infrastructure Network, which is supported by the National Science Foundation (Grant ECS-0335765).

\bibliography{voltmeter}

\begin{thebibliography}{16}
\expandafter\ifx\csname natexlab\endcsname\relax\def\natexlab#1{#1}\fi
\expandafter\ifx\csname bibnamefont\endcsname\relax
  \def\bibnamefont#1{#1}\fi
\expandafter\ifx\csname bibfnamefont\endcsname\relax
  \def\bibfnamefont#1{#1}\fi
\expandafter\ifx\csname citenamefont\endcsname\relax
  \def\citenamefont#1{#1}\fi
\expandafter\ifx\csname url\endcsname\relax
  \def\url#1{\texttt{#1}}\fi
\expandafter\ifx\csname urlprefix\endcsname\relax\def\urlprefix{URL }\fi
\providecommand{\bibinfo}[2]{#2}
\providecommand{\eprint}[2][]{\url{#2}}

\bibitem[{\citenamefont{Besseling et~al.}(2003)\citenamefont{Besseling, Kokubo,
  and Kes}}]{besseling03}
\bibinfo{author}{\bibfnamefont{R.}~\bibnamefont{Besseling}},
  \bibinfo{author}{\bibfnamefont{N.}~\bibnamefont{Kokubo}}, \bibnamefont{and}
  \bibinfo{author}{\bibfnamefont{P.~H.} \bibnamefont{Kes}},
  \bibinfo{journal}{Phys. Rev. Lett.} \textbf{\bibinfo{volume}{91}},
  \bibinfo{pages}{177002} (\bibinfo{year}{2003}).

\bibitem[{\citenamefont{Pruymboom et~al.}(1988)\citenamefont{Pruymboom, Kes,
  van~der Drift, and Radelaar}}]{pruymboom88}
\bibinfo{author}{\bibfnamefont{A.}~\bibnamefont{Pruymboom}},
  \bibinfo{author}{\bibfnamefont{P.~H.} \bibnamefont{Kes}},
  \bibinfo{author}{\bibfnamefont{E.}~\bibnamefont{van~der Drift}},
  \bibnamefont{and} \bibinfo{author}{\bibfnamefont{S.}~\bibnamefont{Radelaar}},
  \bibinfo{journal}{Phys. Rev. Lett.} \textbf{\bibinfo{volume}{60}},
  \bibinfo{pages}{1430} (\bibinfo{year}{1988}).

\bibitem[{\citenamefont{Kokubo et~al.}(2002)\citenamefont{Kokubo, Besseling,
  Vinokur, and Kes}}]{kokubo02}
\bibinfo{author}{\bibfnamefont{N.}~\bibnamefont{Kokubo}},
  \bibinfo{author}{\bibfnamefont{R.}~\bibnamefont{Besseling}},
  \bibinfo{author}{\bibfnamefont{V.}~\bibnamefont{Vinokur}}, \bibnamefont{and}
  \bibinfo{author}{\bibfnamefont{P.~H.} \bibnamefont{Kes}},
  \bibinfo{journal}{Phys. Rev. Lett.} \textbf{\bibinfo{volume}{88}},
  \bibinfo{pages}{247004} (\bibinfo{year}{2002}).

\bibitem[{\citenamefont{Yu et~al.}(2007)\citenamefont{Yu, Heitmann, Song,
  DeFeo, Plourde, Hesselberth, and Kes}}]{yu07}
\bibinfo{author}{\bibfnamefont{K.}~\bibnamefont{Yu}},
  \bibinfo{author}{\bibfnamefont{T.~W.} \bibnamefont{Heitmann}},
  \bibinfo{author}{\bibfnamefont{C.}~\bibnamefont{Song}},
  \bibinfo{author}{\bibfnamefont{M.~P.} \bibnamefont{DeFeo}},
  \bibinfo{author}{\bibfnamefont{B.~L.~T.} \bibnamefont{Plourde}},
  \bibinfo{author}{\bibfnamefont{M.~B.~S.} \bibnamefont{Hesselberth}},
  \bibnamefont{and} \bibinfo{author}{\bibfnamefont{P.~H.} \bibnamefont{Kes}},
  \bibinfo{journal}{Phys. Rev. B} \textbf{\bibinfo{volume}{76}},
  \bibinfo{pages}{220507} (\bibinfo{year}{2007}).

\bibitem[{\citenamefont{Vodolazov and Maksimov}(2000)}]{vodolazov00}
\bibinfo{author}{\bibfnamefont{D.~Y.} \bibnamefont{Vodolazov}}
  \bibnamefont{and} \bibinfo{author}{\bibfnamefont{I.~L.}
  \bibnamefont{Maksimov}}, \bibinfo{journal}{Physica C}
  \textbf{\bibinfo{volume}{349}}, \bibinfo{pages}{125} (\bibinfo{year}{2000}).

\bibitem[{\citenamefont{Plourde et~al.}(2001)\citenamefont{Plourde,
  Van~Harlingen, Vodolazov, Besseling, Hesselberth, and Kes}}]{plourde01}
\bibinfo{author}{\bibfnamefont{B.~L.~T.} \bibnamefont{Plourde}},
  \bibinfo{author}{\bibfnamefont{D.~J.} \bibnamefont{Van~Harlingen}},
  \bibinfo{author}{\bibfnamefont{D.~Y.} \bibnamefont{Vodolazov}},
  \bibinfo{author}{\bibfnamefont{R.}~\bibnamefont{Besseling}},
  \bibinfo{author}{\bibfnamefont{M.~B.~S.} \bibnamefont{Hesselberth}},
  \bibnamefont{and} \bibinfo{author}{\bibfnamefont{P.~H.} \bibnamefont{Kes}},
  \bibinfo{journal}{Phys. Rev. B} \textbf{\bibinfo{volume}{64}},
  \bibinfo{pages}{014503} (\bibinfo{year}{2001}).

\bibitem[{\citenamefont{Benkraouda and Clem}(1998)}]{benkraouda98}
\bibinfo{author}{\bibfnamefont{M.}~\bibnamefont{Benkraouda}} \bibnamefont{and}
  \bibinfo{author}{\bibfnamefont{J.~R.} \bibnamefont{Clem}},
  \bibinfo{journal}{Phys. Rev. B} \textbf{\bibinfo{volume}{58}},
  \bibinfo{pages}{15103} (\bibinfo{year}{1998}).

\bibitem[{\citenamefont{L{\'o}pez et~al.}(1999)\citenamefont{L{\'o}pez, Kwok,
  Safar, Olsson, Petrean, Paulius, and Crabtree}}]{lopez99}
\bibinfo{author}{\bibfnamefont{D.}~\bibnamefont{L{\'o}pez}},
  \bibinfo{author}{\bibfnamefont{W.}~\bibnamefont{Kwok}},
  \bibinfo{author}{\bibfnamefont{H.}~\bibnamefont{Safar}},
  \bibinfo{author}{\bibfnamefont{R.}~\bibnamefont{Olsson}},
  \bibinfo{author}{\bibfnamefont{A.}~\bibnamefont{Petrean}},
  \bibinfo{author}{\bibfnamefont{L.}~\bibnamefont{Paulius}}, \bibnamefont{and}
  \bibinfo{author}{\bibfnamefont{G.}~\bibnamefont{Crabtree}},
  \bibinfo{journal}{Phys. Rev. Lett.} \textbf{\bibinfo{volume}{82}},
  \bibinfo{pages}{1277} (\bibinfo{year}{1999}).

\bibitem[{\citenamefont{Paltiel et~al.}(2000)\citenamefont{Paltiel, Zeldov,
  Myasoedov, Rappaport, Jung, Bhattacharya, Higgins, Xiao, Andrei, Gammel
  et~al.}}]{paltiel00}
\bibinfo{author}{\bibfnamefont{Y.}~\bibnamefont{Paltiel}},
  \bibinfo{author}{\bibfnamefont{E.}~\bibnamefont{Zeldov}},
  \bibinfo{author}{\bibfnamefont{Y.}~\bibnamefont{Myasoedov}},
  \bibinfo{author}{\bibfnamefont{M.~L.} \bibnamefont{Rappaport}},
  \bibinfo{author}{\bibfnamefont{G.}~\bibnamefont{Jung}},
  \bibinfo{author}{\bibfnamefont{S.}~\bibnamefont{Bhattacharya}},
  \bibinfo{author}{\bibfnamefont{M.~J.} \bibnamefont{Higgins}},
  \bibinfo{author}{\bibfnamefont{Z.~L.} \bibnamefont{Xiao}},
  \bibinfo{author}{\bibfnamefont{E.~Y.} \bibnamefont{Andrei}},
  \bibinfo{author}{\bibfnamefont{P.~L.} \bibnamefont{Gammel}},
  \bibnamefont{et~al.}, \bibinfo{journal}{Phys. Rev. Lett.}
  \textbf{\bibinfo{volume}{85}}, \bibinfo{pages}{3712} (\bibinfo{year}{2000}).

\bibitem[{\citenamefont{Babaei~Brojeny and Clem}(2001)}]{babaei01}
\bibinfo{author}{\bibfnamefont{A.~A.} \bibnamefont{Babaei~Brojeny}}
  \bibnamefont{and} \bibinfo{author}{\bibfnamefont{J.~R.} \bibnamefont{Clem}},
  \bibinfo{journal}{Phys. Rev. B} \textbf{\bibinfo{volume}{64}},
  \bibinfo{pages}{184507} (\bibinfo{year}{2001}).

\bibitem[{\citenamefont{Larkin and Ovchinnikov}(1975)}]{larkin75}
\bibinfo{author}{\bibfnamefont{A.~I.} \bibnamefont{Larkin}} \bibnamefont{and}
  \bibinfo{author}{\bibfnamefont{Y.~N.} \bibnamefont{Ovchinnikov}},
  \bibinfo{journal}{Zh. Eksp. Teor. Fiz.} \textbf{\bibinfo{volume}{68}},
  \bibinfo{pages}{1915} (\bibinfo{year}{1975}).

\bibitem[{\citenamefont{Babi{\'c} et~al.}(2004)\citenamefont{Babi{\'c},
  Bentner, S{\"u}rgers, and Strunk}}]{babic04}
\bibinfo{author}{\bibfnamefont{D.}~\bibnamefont{Babi{\'c}}},
  \bibinfo{author}{\bibfnamefont{J.}~\bibnamefont{Bentner}},
  \bibinfo{author}{\bibfnamefont{C.}~\bibnamefont{S{\"u}rgers}},
  \bibnamefont{and} \bibinfo{author}{\bibfnamefont{C.}~\bibnamefont{Strunk}},
  \bibinfo{journal}{Phys. Rev. B} \textbf{\bibinfo{volume}{69}},
  \bibinfo{pages}{92510} (\bibinfo{year}{2004}).

\bibitem[{\citenamefont{Clarke}(1966)}]{clarke66}
\bibinfo{author}{\bibfnamefont{J.}~\bibnamefont{Clarke}},
  \bibinfo{journal}{Philosophical Magazine} \textbf{\bibinfo{volume}{13}},
  \bibinfo{pages}{115} (\bibinfo{year}{1966}).

\bibitem[{\citenamefont{Gammel et~al.}(1991)\citenamefont{Gammel, Schneemeyer,
  and Bishop}}]{gammel91}
\bibinfo{author}{\bibfnamefont{P.~L.} \bibnamefont{Gammel}},
  \bibinfo{author}{\bibfnamefont{L.~F.} \bibnamefont{Schneemeyer}},
  \bibnamefont{and} \bibinfo{author}{\bibfnamefont{D.~J.}
  \bibnamefont{Bishop}}, \bibinfo{journal}{Phys. Rev. Lett.}
  \textbf{\bibinfo{volume}{66}}, \bibinfo{pages}{953} (\bibinfo{year}{1991}).

\bibitem[{\citenamefont{Safar et~al.}(1992)\citenamefont{Safar, Gammel, Bishop,
  Mitzi, and Kapitulnik}}]{safar92}
\bibinfo{author}{\bibfnamefont{H.}~\bibnamefont{Safar}},
  \bibinfo{author}{\bibfnamefont{P.~L.} \bibnamefont{Gammel}},
  \bibinfo{author}{\bibfnamefont{D.~J.} \bibnamefont{Bishop}},
  \bibinfo{author}{\bibfnamefont{D.~B.} \bibnamefont{Mitzi}}, \bibnamefont{and}
  \bibinfo{author}{\bibfnamefont{A.}~\bibnamefont{Kapitulnik}},
  \bibinfo{journal}{Phys. Rev. Lett.} \textbf{\bibinfo{volume}{68}},
  \bibinfo{pages}{2672} (\bibinfo{year}{1992}).

\bibitem[{\citenamefont{Clem and Sanchez}(1994)}]{Clem94}
\bibinfo{author}{\bibfnamefont{J.~R.} \bibnamefont{Clem}} \bibnamefont{and}
  \bibinfo{author}{\bibfnamefont{A.}~\bibnamefont{Sanchez}},
  \bibinfo{journal}{Phys. Rev. B} \textbf{\bibinfo{volume}{50}},
  \bibinfo{pages}{9355} (\bibinfo{year}{1994}).

\end{thebibliography}
\end{document}